\documentstyle[sprocl,amssymb,amsbsy,epsf]{article}
\title{A Numerical Approach to Binary Black Hole Coalescence}
\author{Lee Samuel Finn}
\address{Northwestern University, Physics and Astronomy Department,
Evanston Illinois 60208-3112, USA}
\begin{document}
\maketitle \abstracts{ The nature of binary black hole coalescence is
the final, uncharted frontier of the relativistic Kepler problem. In
the United States, binary black hole coalescence has been identified
as a computational ``Grand Challenge'' whose solution is the object of
a coordinated effort, just reaching its half-way point, by more than
two-score researchers at nearly a dozen institutions. In this report I
highlight what I see as the most serious problems standing between us
and a general computational solution to the problem of binary black
hole coalescence: 
\begin{itemize}
\item the computational burden associated of the problem
based on reasonable extrapolations of present-day
computing algorithms and near-term hardware developments; 
\item some of the computational issues associated with those
estimates, and how, through the use of different or more sophisticated
computational algorithms we might reduce the expected burden; and
\item some of the physical problems associated with the development of
a numerical solution of the field equations for a binary black hole
system, with particular attention to work going on in, or in
association with, the Grand Challenge.
\end{itemize}
}

\section{Introduction}

In its simplest form, the relativistic Kepler problem involves the
orbital evolution of two black holes in an asymptotically flat
universe with no-incoming-radiation boundary conditions. Consider a
binary black hole system at early times when we can speak, in the
sense of the correspondence principle, of large orbital angular
momentum and orbital energy close to zero. The evolution of such a
system can be regarded in three parts:
\begin{enumerate}
\item At early times the evolution is adiabatic: the black holes
circle each other and the orbits change on timescales long compared to
the orbital period.
\item As the separations become smaller the orbit evolves more quickly
and the adiabatic approximation breaks down. Shortly thereafter the
black holes plunge together and coalesce to form a single, large and
highly dynamical black hole.
\item Finally, the descendant black hole radiates away its
perturbation and settles down to a stationary state that evolves no
further.
\end{enumerate}

The intermediate period of binary evolution described above, beginning
when the binary separation is on order $13M$ for equal mass black
holes and ending with the emergence of a single perturbed black hole,
is of great interest to three communities of scientists:
\begin{itemize}
\item to the relativist it represents the missing link in the solution
of relativistic Kepler problem, a regime both strongly non-linear and
highly dynamical, without whose understanding the initial binary
cannot be matched to the final stationary black hole spacetime;
\item to the astrophysicist it represents a period of strong
gravitational-wave radiation, whose detection may yield the most
direct evidence for the existence of black holes, a new determination
of the Hubble Constant, and clues to the density and mass distribution
of black holes in the universe;
\item finally, to the computational physicist it represents a complex
physical system whose numerical solution exceeds the capabilities of
the present day combination of computing algorithms and hardware and
will challenge those of the next generation.
\end{itemize}

Both the initial period of adiabatic evolution and the final era of
black hole ``ring-down'' can be understood quantitatively using
perturbative techniques. The details of the intermediate period ---
when the evolution is non-adiabatic and the black holes coalesce ---
is not understood quantitatively and does not appear amenable to
approximate treatment short of a fully numerical solution.

In view of the significant computational challenges and great
scientific interest in the solution to this outstanding problem of
physics, the United States National Science Foundation has funded a
``Computational Grand Challenge Team'' with the goal of developing, in
five years, a general, extensible computational solution to the
problem of binary black hole coalescence. The leading investigators of
that team are
\begin{itemize}
\item Richard Matzner and James Brown (University of Texas, Austin);
\item Stuart Shapiro (Center for Astrophysics and Relativity);
\item Charles R. Evans and James York (University of North Carolina,
Chapel Hill).
\item Saul Teukolsky (Cornell University);
\item Faisal Saied, Paul Saylor, Edward Seidel, Larry Smarr
(National Center for Supercomputing Applications and University of
Illinois, Champaign-Urbana);
\item Lee Samuel Finn (Northwestern University);
\item Pablo Laguna (Pennsylvania State University);
\item Jeffrey Winicour (Pittsburgh University); and
\item Geoffrey Fox (Syracuse University);
\end{itemize}
In addition, close and fruitful collaborations have been struck
between members of this Grand Challenge Team, Wai-Mo Suen
(Washington University, St. Louis) and Nigel Bishop (University of
South Africa).

In this report I highlight what I see as the most serious problems
standing between us and a general computational solution to the
problem of binary black hole coalescence. In doing so, I hope to
convey a real sense of where we are and, more importantly, where we
are heading. In \S\ref{sec:ivp} I give a conceptual overview of
how we formulate the evolution of a vacuum spacetime as an initial
value problem. With this as background, in \S\ref{sec:burden} I
discuss the computational burden associated with binary black hole
coalescence, based on reasonable extrapolations of present-day
computing algorithms and near-term hardware developments. In
\S\ref{sec:algorithms} I discuss some of the computational issues
associated with those estimates, and how, through the use of different
or more sophisticated computational algorithms we might reduce the
expected burden. Finally, in \S\ref{sec:physics} I survey some of the
physical problems associated with the numerical solution of the field
equations for a binary black hole system, focusing on work going on in
association with the Grand Challenge.

\section{Spacetime as an initial value problem\label{sec:ivp}}

Our problem, simply stated, is to evolve forward in time an initial,
spacelike hypersurface with two black holes. As posed, the problem
distinguishes naturally between space (defined by the initial
hypersurface) and time (generally orthogonal to the initial value
hypersurface). In the conventional numerical treatment of initial
value problems we choose a coordinate system that enforces a
distinction between space and time everywhere: we foliate spacetime
with spacelike hypersurfaces, of which the earliest is the initial
value hypersurface, describe spacetime's geometry in terms of the
intrinsic and extrinsic geometry of these hypersurfaces, and formulate
the field equations to allow us, given the intrinsic and extrinsic
geometry of one spacelike hypersurface, to find the same for the
next. The collection of spacelike hypersurfaces together with their
intrinsic and extrinsic geometry completely determine the geometry of
a four-dimensional volume of spacetime; in our case, the spacetime of
a binary black hole system.

The basic formulation of this $3+1$ split of spacetime is due to
Arnowitt, Deser and Misner~\cite{arnowitt62a}. The spacelike
hypersurfaces of the foliation are level surfaces of coordinate time,
and three coordinate functions are defined to label points on each
hypersurface.  There is some ambiguity in the choice of spatial and
time coordinates corresponding to the gauge freedom of the theory:
this freedom allows us to choose the relation between spatial
coordinates on successive hypersurfaces ({\em i.e.,} lines of constant
spatial coordinate may be shifted relative to the hypersurface
orthogonal) and also the elapsed proper time along the hypersurface
orthogonal joining successive slices.

\begin{figure}
\epsfxsize=\columnwidth\epsffile{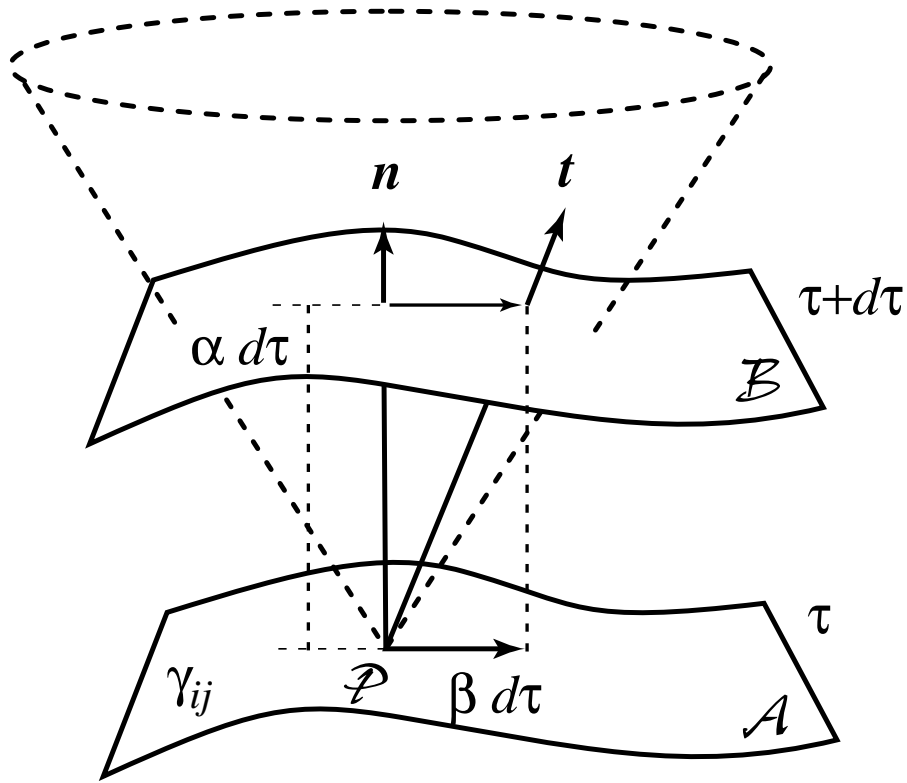}
\caption{In numerical relativity, initial value problems are treated
by introducing a coordinate system that distinguishes between space
and time; the field equations are then formulated in a way that permits
the intrinsic and extrinsic geometry of one slice (the initial data)
to determine the same on successive slices. Here is shown the
relationship between the spacelike hypersurfaces, which are level
surfaces of the coordinate time, their normals, and the spatial
coordinates on successive hypersurfaces in the conventional $3+1$
decomposition of spacetime. See \S\protect{\ref{sec:ivp}} for
details.}\label{fig:slicing}
\end{figure}

Figure \ref{fig:slicing} (adapted from York~\cite{york79a})
illustrates how two successive hypersurfaces of the spacetime
foliation are related to each other. In the coordinate system used in
the numerical evolution each of these surfaces is a slice of constant
coordinate time; the coordinate time difference between the two slices
shown is $d\tau$. Focus attention on the bottom slice (${\cal A}$),
which is the level surface corresponding to coordinate time
$\tau$. The intrinsic geometry of this slice is described by the
3-metric $\gamma_{ij}$, which is the projection of the spacetime
4-metric on ${\cal A}$. The tangent to the hypersurface normal at
point ${\cal P}$ (on $\cal A$) is $\bf n$, and the tangent to the
world line of a coordinate-stationary observer at ${\cal P}$ is $\bf
t$.

The upper hypersurface (${\cal B}$) is the spacelike 3-surface of
constant coordinate time $\tau+d\tau$. The elapsed proper time between
${\cal A}$ and ${\cal B}$ along $\bf n$ is $\alpha({\cal P})\,d\tau$,
where $\alpha$ is the {\em lapse function.} The coordinate (gauge)
freedom of the theory gives us some freedom in the choice of the
lapse. Lying in each spacelike hypersurface is a 3-vector field
$\beta$, called the {\em shift}, that describes how worldlines of
coordinate stationary observers deviate from the hypersurface
orthogonal $\bf n$. The choice of three-vector $\beta$ exhausts our
coordinate freedom. An exceptionally readable description of the $3+1$
decomposition of spacetime is given in York~\cite{york79a}.

Having chosen the $3+1$ coordinates, it is necessary to define fields
on each hypersurface from which the hypersurface spatial metric and
extrinsic curvature can be determined. The Einstein field equations
can then be cast in terms of those variables, leading to constraint
equations, which do not involve time derivatives of the fields, and
evolution equations, which do. Constraint equations must be satisfied
by the fields on each hypersurface, while evolution equations involve
reference to future surfaces of the foliation and are used to
determine the fields on successive hypersurfaces.  The identification
of field variables, formulation of their equations of motion, and the
specification of boundary conditions and initial data are all areas of
active research today and are discussed in \S\ref{sec:physics}.

\section{Computational Burden\label{sec:burden}}

Should we tomorrow find ourselves suddenly in command of unlimited
computing resources we would be no closer to a numerical solution to
binary black hole coalescence than we are today: there remain open
questions --- regarding the appropriate choice of field variables
(and, correspondingly, the equations to be solved), the numerical
method of solution and the interpretation of both initial data and the
results --- that must be settled before we can construct a numerical
code that ``solves'' the black hole binary coalescence
problem. Nevertheless, it is instructive to anticipate the ultimate
computational burden of a binary black hole coalescence code using
present-day techniques.

\subsection{Memory}

As discussed in \S\ref{sec:ivp}, the numerical model of binary black
hole coalescence is constructed from an initial, spacelike,
three-dimensional slice of spacetime, boundary conditions, and the
Einstein field equations. We cannot evolve a slice of infinite volume
(nor would we want to if we could); so, we choose the smallest region
of spacetime that includes the system of interest and on which we can
impose sensible boundary conditions. All simulations proceed by
carrying approximate values of field quantities at isolated points
within the volume, and the variations of these fields on the important
length scales ({\em e.g.,} black hole mass, radiation wavelength) must
be suitably resolved. Together, these constraints determine the
primary memory that must be devoted to each timestep in the
simulation.

Consider a system of two roughly equal black mass holes with total
mass $M$ in a circular orbit of radius $\sim6M$. The quadrupole order
radiation wavelength is on order $100M$, which is substantially
greater than the system's size.  To pose sensible outer boundary
conditions and extract the radiation from the system we need the outer
boundary of the computational volume at least one wavelength away from
the from the system's center~\cite{abrahams90a}; thus, the
computational volume is $\gtrsim10^6M^3$. The simulation resolution
must be great enough to resolve the individual black hole and the
radiation; a resolution of $\Delta x\lesssim M/20$ is a reasonable
assumption. Assuming uniform resolution throughout each spatial slice
requires that the field variables be maintained at $\gtrsim10^{10}$
isolated points within the computational volume.

On each slice of constant time are equations governing the evolution
of the slice's spatial metric and its extrinsic curvature. In addition
to these fields are several auxiliary quantities that appear
frequently enough in the evolution and constraint equations that it is
sensible to compute them once on each slice and re-use their stored
values when necessary. Thus, there are on order 50 field and auxiliary
quantities that must be stored, in double precision, at each of the
$\gtrsim10^{10}$ points on a given hypersurface, leading to a final
estimate of $\gtrsim4\times10^{12}$~bytes of primary memory for the
computational volume. An equivalent amount of secondary storage ({\em
e.g.,} disk storage) is required {\em for each} slice of a time
evolution that we wish to store for later examination.

\subsection{Execution speed}

A complete simulation involves the evolution of the computational
volume from an initial data set of two separate black holes, through
their coalescence, to a final data set consisting of a single
perturbed black hole. The simulation's spatial resolution, together
with the numerical algorithms used, determines its temporal
resolution. The simulations duration and the required temporal
resolution determine the number of intermediate spatial slices that
must be computed in passing from the initial to the final
hypersurface. If we use finite difference methods (cf.~\S\ref{sec:fd})
for solving the partial differential equations that evolve each slice
to the next, then the computational work per slice (measured in
floating point operations) depends linearly on the slice's volume and
inversely on its spatial resolution. The work per slice and the number
of slices then yields the total number of floating point operations
required for the full simulation. Insisting that such a simulation
take a reasonable amount of time to perform leads to a minimum
requirement for the computing speed in floating point operations per
second, or flop/s.

Assume that the simulation starts with black holes in a nearly circular
orbit at a separation of $\sim8M$, and that there are at most two
orbits --- corresponding to an elapsed time of $\sim250M$ --- before
coalescence.  After coalescence the final black hole spacetime will be
highly perturbed and the simulation will need to run several (say,
five) black hole fundamental mode periods before the perturbations are
small enough that the numerical simulation can be matched to a
perturbation calculation. The total simulated time is then
$T\gtrsim500M$. If the field variable characteristics are light cones
then the simulation's time resolution is on order its spatial
resolution, or $\lesssim M/20$, and a complete simulation requires
$\gtrsim10^4$ timesteps. If the size and spatial resolution of each
slice require that field and auxiliary variables be maintained at
$10^{10}$ points at each timestep, then there are $10^{14}$ updates
(of field and auxiliary variables) that occur in the course of a
complete simulation. Modern three-dimensional codes require
approximately $5\times10^3$ flop/s per update~\cite{anninos95c};
consequently, there are $\gtrsim5\times10^{17}$ floating operations in
a simulation. If we require that each simulation take less than
$\sim12$~hours of computer time, then the computer must be capable of
$\gtrsim10^{13}$~flop/s, or $10$~Tflop/s (teraflop/s).

\subsection{Conclusions}

What do these numbers mean?

At this writing (November 1995), the ``fastest'' general-purpose
supercomputer in the world (as measured by its performance on the {\sc
Linpack} benchmark~\cite{dongarra94a}) is the Fujitsu ``Numerical Wind
Tunnel'', installed at the National Aeronautics Laboratory in
Japan. Its realized peak performance is $170\times10^9$~flop/s, or
170~Gflop/s, compared to a ``theoretical'' peak performance of
236~Gflop/s~\footnote{A computer's theoretical peak performance may be
interpreted as that performance it is guaranteed {\em never} to
exceed.} Of those computers generally available for academic research,
the fastest is the 512 node IBM SP2, installed at the Cornell Theory
Center, with a realized peak performance of 88.4~Gflop/s and a
theoretical peak performance of 136~Gflop/s. The National Center for
Supercomputing Applications (NCSA) numerical relativity group reports
a realized speed of 14.6~Gflop/s on a Thinking Machines CM-5 (whose
theoretical peak performance is 63~Gflop/s). If we assume this latter
ratio to be typical of the realized to theoretical performance for the
next generation parallel-processing computers, then we anticipate the
need for a two order of magnitude increase in the computing
performance, on computers with terabyte primary storage, before
generic three-dimensional binary black hole coalescence calculations,
made using current methodology, are tractable.

\section{Computing algorithms\label{sec:algorithms}}

The conclusion of the previous section --- that a two
order-of-magnitude increase in available computing speed is required
to make binary black hole coalescence calculations tractable --- comes
accompanied by the important caveat that no change in either the
computational algorithms or problem formulation affects significantly
the assumptions that underly it.

The estimate of the computing burden of a generic binary black hole
coalescence calculation is most sensitive to the simulation's spatial
resolution: if, without loss of accuracy or significant increase in
the cost of the computation per update, the spatial resolution can be
reduced by a factor of two in each dimension, the memory burden is
reduced by a factor of 8 and the required computational speed by a
factor of 16.  Algorithm changes can affect the attainable accuracy as
a function of the resolution; so, as we hope for faster computers we
not neglect the search for more efficient (in the sense accuracy for a
given resolution) computing algorithms. In this section I describe two
different methods for solving partial differential equations on
computers and comment on the strengths, weaknesses, and promise of
each.

\subsection{Numerical solution of partial differential equations}

Computers don't do calculus: differential equations are solved
numerically by converting them to a set of algebraic equations that
are solved by the more elementary operations of addition, subtraction,
multiplication and division. There is a range of methods used to
approximate partial differential equations by algebraic ones; however,
the practice of numerical relativity has focused almost exclusively on
the {\em finite difference} approximation.

\subsubsection{Finite differencing\label{sec:fd}}

In a finite difference approximation the derivatives in the
differential equations are replaced by approximations formed from
field variable values at discrete points. For example, an
approximation to the second derivative of $u$ might be written as 
\begin{equation}
{d^2 u\over dx^2}(x_i) \rightarrow 
{u(x_{i+1}) -2u(x_i)+u(x_{i-1}) \over (\Delta x)^2},
\end{equation}
where the $x_i$ are discrete points and $\Delta
x=x_{i+1}-x_i$. Clearly the points $x_i$ do not need to be spaced
equally; similarly, the algebraic approximations for different
derivatives can be made arbitrarily complex and accurate to high order
in the spacing. With a choice of points and a ``differencing scheme''
for converting derivatives to ratios of finite differences, any set of
differential equations can be converted to a set of algebraic
equations that can be solved using standard methods of linear algebra.

In a finite difference approximation, the solution to the approximate
equations converges upon the solution to the exact equations as a
finite power of the resolution $h\simeq|x_i-x_{i-1}|$. The
exponent is called the order of the differencing scheme; for example,
\begin{equation}
{\left[u(x_{i+1})-u(x_i)\right](x_{i}-x_{i-1})
\over (x_{i+1}-x_i)(x_{i+1}+x_{i-1})}
+
{\left[u(x_i)-u(x_{i-1})\right](x_{i+1}-x_i)
\over (x_{i}-x_{i-1})(x_{i+1}+x_{i-1})} 
=  {d u\over dx}(x_i)+{\cal O}(h^2)
\end{equation}
is a second-order difference approximation to $du/dx$ at $x_i$ for
unevenly spaced points $x_i$.  For problems with smooth solutions,
the difference between the solutions to the differential and
difference equations is, for small $h$, proportional to $h^N$, where
$N$ is related to the order of the least accurate differencing scheme
used to approximate the system and its boundary conditions. 

Finite difference approximations are relatively simple and
straightforward to program. They are also flexible: the choice of grid
points and their spacing can conform to irregular boundaries or be
concentrated in areas where increased resolution is needed.
On the other hand, finite difference methods are sensitive to
coordinate singularities: accurate difference operators for points on
or near the origin or polar axis of a spherical coordinate system are
difficult to construct and generally. 
The implementation of boundary conditions that are not algebraic in
the evolving fields ({\em e.g.,} a Sommerfeld out-going wave boundary
condition) often involves introducing grid points ``beyond the
boundary'' that are not evolved, but are reset at each timestep so
that the differencing scheme, applied at the boundary, leads to an
appropriate approximation.
Finally, there are a variety of finite difference schemes, whose
errors scale with resolution identically but whose performance, at
boundaries and coordinate singularities, differ greatly. An acceptable
choice of differencing scheme near these singularities is generally
problem specific and requires fine-tuning.

The large computational burden of the binary black hole coalescence
calculation is due primarily to the high spatial resolution required
in each spacelike slice. High resolution is not required
everywhere in the slice, however, but only where the field variables
are changing on short lengthscales. Near the black holes, resolution
on sub-$M$ scales is necessary to resolve the spacetime curvature
accurately; however, far from the holes (in the larger part of the
computational volume) the shortest lengthscale is a radiation
wavelength and a much coarser resolution will provide the same level
of accuracy. If we give up the convenience of uniform resolution,
the computational burden of a finite difference calculation can be
reduced dramatically by allowing the spatial resolution to vary so
that the approximation is equally accurate everywhere.

This is the idea behind {\em adaptive mesh refinement:} in any region
resolve only so far as is necessary to attain the desired local
accuracy. Adaptive mesh refinement is being pursued aggressively
within the collaboration by physicists and computer scientists at the
University of Texas and Syracuse University. In the present example,
high ($M/20$) resolution is needed only near the black holes: for most
of the volume of each spatial slice resolution of order $M$ to $10M$
may be more appropriate. Adaptive mesh refinement might then lead to a
reduction in the memory required per spatial slice of a factor $10^4$
over a monolithic grid, and a reduction in the speed requirements by a
factor $10^5$ (less the increased cost per time step).

\subsubsection{Collocation pseudo-spectral methods}

In finite differencing the differential equations are approximated by
algebraic ones that are then solved numerically. A complementary
approach approximates not the equations, but their solution. In a
{\em collocation pseudo-spectral approximation,} the solution $u$
is written as a sum over a set of basis functions $\phi_n$ with
unknown coefficients:
\begin{equation}
u(x) \simeq \sum_{n=0}^N c_n \phi_n(x).
\end{equation}
In two or more dimensions it is typical to choose a separate basis in
each dimension and express $u({\bf x})$ as a sum over products of
the basis functions in each dimension: {\em e.g.,}
\begin{equation}
u(x,y) \simeq \sum_{m,n=0}^{M,N} c_{mn} \phi_m(x)\psi_n(y).
\end{equation}
Algebraic equations for the $N$ coefficients $c_n$ of the approximate
solution are found by insisting that the approximate solution satisfy
exactly the equations and boundary conditions at an equal number of
{\em collocation points} $x_i$. The basis functions and collocation
points are typically chosen to exploit a discrete orthogonality
relation: for example, the basis might be a finite subset of Chebyshev
polynomials $\phi_n=T_n$ on the domain $x\in[-1XSAS,1]$ and the collocation
points $x_k=\cos(\pi k/N)$ so that
\begin{equation}
\sum_{k=1}^{N} T_i(x_k)T_j(x_k) = 
\left\{
\begin{array}{ll}
0&i\neq j\\
N/2&i=j\neq0\\
N&i=j=0
\end{array}
\right.
\end{equation}

For problems with $C_\infty$ solutions a collocation pseudo-spectral
approximation is extremely efficient: with a suitable basis the
asymptotic rate of convergence of the approximate to the real solution
is exponential~\cite{gottleib77a}, which is faster than can be
achieved with any finite difference approximation. The choice of
appropriate basis is not difficult, either: a Fourier decomposition in
variables where periodic boundary conditions hold and a Chebyshev
decomposition elsewhere is usually sufficient. The computation cost
per coefficient (or collocation point) grows more rapidly than
linearly; however, for Fourier and Chebyshev expansion bases that
growth is no faster than $N\log N$. In the asymptotic regime, then,
the marginal return of accuracy on an investment of computing
resources is greater for a solution found using a collocation
pseudo-spectral approximation than for one found using a finite
difference approximation.

Larry Kidder and I have been investigating the use of collocation
pseudo-spectral methods in numerical relativity applications. Our
first investigations have focused on solving the initial value problem
for a single black hole with Brill waves, a problem also studied using
finite difference methods.~\cite{bernstein94a} 
On typical problems we find exponential convergence of the solution
with spectral bases involving greater than 5 angular and 5 radial
Chebyshev polynomials.
Comparing spectral and finite difference methods on the same problem,
a spectral expansion using 20 radial and 20 angular basis functions
(400 collocation points) achieved the same level of accuracy as a
finite difference calculations using 400 radial and 105 angular
subdivisions ($4\times10^4$ grid points); furthermore, the accuracy of
the spectral calculation could be increased to machine precision (8
orders of magnitude better than the best finite difference calculation
reported~\cite{bernstein94a}) with a basis of 70 radial and 70 angular
(Chebyshev) functions.\cite{kidder-p1}
Finally, while the finite difference calculations were performed on
the NCSA Cray Y-MP and Cray 2, the $20\times20$ spectral solution took
30~s, and the most highly resolved spectral calculation ($70\times70$)
less than 10~minutes, on a Sun SPARCstation II.

Investigation of collocation pseudo-spectral methods for vacuum
gravity have only just begun and their ultimate usefulness remains to
be demonstrated. Despite their great promise, however, they are {\em
not} a useful tool for this Grand Challenge application, which is
committed to having a working solution by the end of 1998. To meet
this goal, the collaboration must focus its efforts on increasing the
efficiency of solutions based on the more mature (in this application)
finite difference approximation.

\section{Issues of physics\label{sec:physics}}

\subsection{Radiation and the outer boundary\label{sec:outerbc}}

A principal goal of the Grand Challenge project is to determine the
gravitational radiation arising from the coalescence of a binary black
hole system. The appropriate boundary conditions (no incoming
radiation) are most naturally posed at past null infinity and the
radiation is naturally identified at future null infinity; however,
the initial data for the numerical evolution is a spacelike
hypersurface of finite volume. Consequently, boundary conditions must
be posed and the radiative component of the fields identified at the
boundary of this slice --- at finite distance from the binary system
and not at past and future null infinity. For the outer boundary of
the spacelike slice, then, we must formulate boundary conditions to be
applied at a finite distance from the origin that, as nearly as
possible,
\begin{enumerate}
\item do not introduce any spurious radiation (either as incoming
radiation from past null infinity or as reflections from the boundary)
into the computational volume, and
\item allow us to determine the radiation waveforms, radiated power,
{\em etc.,} from the interior fields.
\end{enumerate}

Two approaches to this outer boundary condition problem --- which have
come to be referred to as ``radiation extraction'' and
``Cauchy-characteristic matching'' --- are under investigation. 
In radiation extraction it is assumed that the spacetime exterior to
the worldtube enclosing the computational volume is well-approximated
by a perturbation of an exact, stationary spacetime ({\em e.g.,}
Minkowski, Schwarzschild or Kerr spacetimes). Using the appropriate
perturbation equations ({\em e.g.,} the Zerilli equation for an
approximately Schwarzschild exterior) the metric perturbations
determined at the edge of the computational volume are readily
``propagated'' to large distances where the asymptotic radiation
fields (in, {\em e.g.,} TT gauge), radiated power, {\em etc.,} can be
determined~\cite{abrahams88a,abrahams90a,price94a,abrahams95a}. At the
same time, boundary conditions corresponding to no incoming radiation
from the exterior spacetime are imposed on the boundary of the
interior volume.

\begin{figure}
\epsfxsize=\columnwidth\epsffile{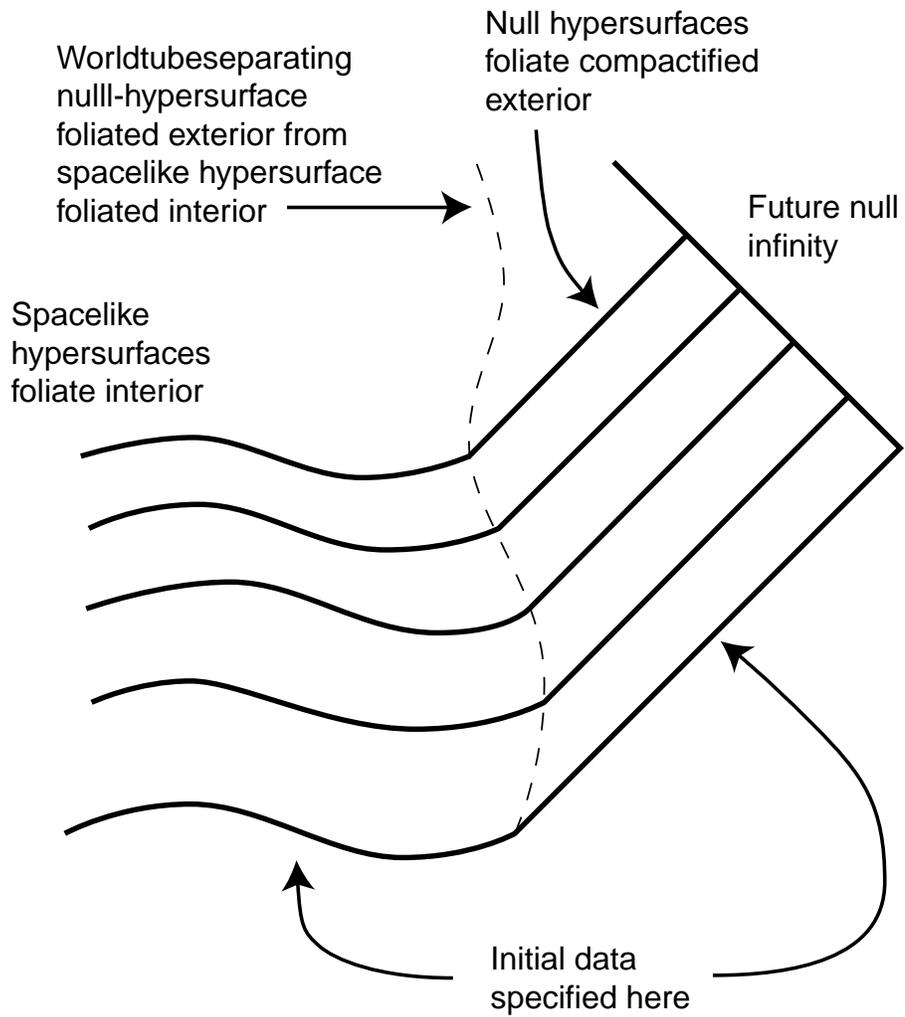}
\caption{In Cauchy-characteristic matching each finite volume
spacelike slice is joined to a null hypersurface that extends to
future null infinity where the outgoing radiation is readily
determined. See \S\protect{\ref{sec:outerbc}} for more
details.}\label{fig:cc}
\end{figure}

In radiation extraction the worldtube boundary of the evolving
spacelike hypersurfaces are assumed to match onto a perturbation of an
exact, stationary spacetime. In the Cauchy-characteristic matching
method (under development at Pittsburgh and Penn State) the worldtube
is joined to a foliation of the exterior spacetime by null
hypersurfaces that extend to future null infinity. The generators of
these null hypersurfaces are the future-directed null geodesics
orthogonal to the worldtube boundary of the spacelike-hypersurface
foliation. By choosing a compactified radial coordinate on the null
hypersurfaces, future null infinity becomes a sphere at finite radial
coordinate (see figure \ref{fig:cc}). The spacelike interior foliation
and the null exterior foliation are evolved together using the field
equations; in this way, radiation generated in the interior that
reaches the worldtube boundary appears immediately at future null
infinity.

Radiation extraction and Cauchy-characteristic matching can be
compared in several ways. Both methods attempt to impose only
outgoing-wave boundary conditions (a strict implementation of
no-incoming radiation boundary conditions requires the ability to
control past null infinity, which involves the entire past history of
the initial data slice). Radiation extraction is simpler to implement
--- it involves only perturbations about flat, Schwarzschild or Kerr
spacetimes --- while Cauchy-characteristic matching involves the field
equations in all their partial-differential glory, along with a
complicated matching between the Cauchy and characteristic evolution
codes. On the other hand, perturbative treatments are only
satisfactory when the fields are perturbative, while the
Cauchy-characteristic method allows boundary conditions on the spatial
slice to be posed where the geometry is strongly dynamical (as long as
the exterior fields are never so axisymmetric that the hypersurface
generators cross); additionally, the boundary conditions come from
future null infinity and not a finite distance approximation.
Finally, while Cauchy-characteristic matching has been demonstrated
for non-trivial vacuum gravity simulations~\cite{clarke95a,dubal95a},
radiation extraction is the more mature
technology~\cite{abrahams88a,abrahams90a,abrahams95a}.

\subsection{Apparent Horizon Boundary Conditions\label{sec:ahbc}}

A numerical computation deals only with finite quantities;
consequently, the spatial computational volume must avoid approaching
too closely any spacetime singularities. In evolving a black hole
spacetime it has been conventional to choose an initial slice that
passes through the black hole throat, thus avoiding the
singularity. As the initial slice is evolved, the lapse coordinate
gauge freedom is exploited to ``freeze'' the evolution at and near
throat, thus avoiding the interception of the singularity.

There is a price paid in avoiding the singularity in this manner,
however: as time progresses the volume of the spatial slice grows
exponentially near the throat as it is stretched to connect the
evolving space far from the black hole to the unevolved space close to
the throat. With the exponential stretching comes an unavoidable
exponential growth in the computational resources devoted to the
physically uninteresting transition between the evolved and unevolved
regions. The net result is that, even if all other numerical pitfalls
are overcome, the inability to resolve the stretching throat destroys
the accuracy of the simulation more rapidly than a single orbital
period of a binary black hole system.

The rapid change of the metric and extrinsic curvature near the throat
reflect the singularity avoiding gauge choice and not anything of
physical interest. Furthermore, because this region lies inside the
event horizon it can have no physical effect on the spacetime outside
the horizon.  This suggests a different approach to the evolution of a
the exterior black hole spacetime: use the horizon as a boundary,
posing down-going radiation boundary conditions on it. Of course, the
event horizon is not suitable for this purpose since it is known only
after the development of the spacetime is complete; however, apparent
horizons, even though gauge dependent, have all the requisite
properties and can be identified on a spatial slice.

Apparent horizon boundary conditions have been pioneered by the
Washington University/NCSA~\cite{seidel92b,anninos95b} and the
Cornell/Center for Astrophysics and Relativity~\cite{scheel95a} groups.

\subsection{Gauge and physics}

Having chosen a spacetime slicing, the apparent horizon is a causal
surface in the sense that nothing {\em physical} propagates outward
from it. It is this property that makes it possible to conceive of
evolving the volume outside the horizon without reference to the
volume inside, as discussed in
\S\ref{sec:ahbc}.  While physical effects propagate at the speed of
light (or less for ``tail'' terms or simulations involving matter
fields), non-physical (gauge) fields can propagate
superluminally. {\em If we are to use the apparent horizon as a
boundary surface we must be certain that none of the characteristics
of the (gauge dependent) fields we are evolving are superluminal.}

The conventional formulation of the Einstein field equations in the
$3+1$ spacetime decomposition determines equations of motion for the
spatial metric ($\gamma_{ij}$) and the intrinsic curvature of the
hypersurface ($K_{ij}$) of each spacelike hypersurface. These coupled,
non-linear differential equations are not hyperbolic; consequently, in
this formulation there is no rigorous domain of dependence or region
of influence for the fields at a point in spacetime and no assurance
that the (gauge dependent) fields $\gamma_{ij}$ and $K_{ij}$ at a
point outside an apparent horizon do not depend on the fields inside
the horizon.

Recent work, carried out independently by several different
groups~\cite{bona95a,abrahams95b,vanputten95Pa} but motivated by its
signifcance for numerical relativity calculations, has led to a number
of new formulations of the Einstein field equations that are
explicitly hyperbolic. In the most general of these
formulations~\cite{bruhat95Pa,abrahams95a} the equations of motion for
the fields (now the extrinsic curvature and its Lie derivative along
the time axis $\bf t$, cf.~fig.~\ref{fig:slicing}) are hyperbolic {\em
independent} of the choice of spatial gauge (shift vector $\beta$),
and the only non-zero characteristic speed is $c$.

While the significance of this hyperbolic formulation of the Einstein
field equations extends far beyond it numerical relativity
application, its importance to numerical relativity is difficult to
overstate.
Since, in this formulation, the characteristics are well-defined and
light-like, apparent horizon boundary conditions can now be rigorously
and confidently imposed in black hole spacetimes.
Because all fields are either Lie-dragged along the time axis or
propagate along light cones, imposition of boundary conditions on the
outer-boundary of a spacelike slice (or matching to an exterior Cauchy
evolution) is greatly simplified.
As might be expected (since the only non-zero propagation speed is
$c$), the perturbative reduction of the field equations lead directly
to gauge-invariant quantities, simplifying greatly radiation
extraction.
Finally, a hyperbolic formulation of the field equations gives
numerical relativity direct access to an extensive, but hitherto
unusable, body of mature work on the numerical evolution of
hyperbolic systems.

\subsection{Assembling the pieces}

As discussed in the introduction, binary black hole coalescence is
only one part of a larger problem --- the relativistic Kepler
problem. In the context of this larger problem, a numerical solution
to black hole coalescence takes as its initial conditions the results
of a perturbative treatment of binary inspiral, and its end-results
become the initial conditions for a perturbative treatment of black
hole ringdown. In this way an initial black hole binary system, whose
components are so widely separated that they can be regarded as
isolated, is associated with a final black hole of given mass and
angular momentum (and also gravitational radiation from the inspiral,
coalescence and ringdown).

When the binary component separation is large then one can speak sensibly
of the component masses and spins as well as the system's orbital
energy and angular momentum. When the component separation is
small, however, these familiar quantities are no longer available to
us: as the orbit tightens the distinctions between the binary
component masses and the system's total mass, or between the spin and
orbital angular momentum, or between the total mass, the orbital
energy, and the energy in the radiation, become blurred and lose their
meaning. 

This is as it should be --- the final state is, after all, a quiescent
black hole characterized only by its mass and angular momentum;
nevertheless, it complicates our ability to place the initial data in
an astrophysical setting.  As discussed in \S\ref{sec:burden}, the
combination of high computational burden and limited resources will
require that we start each binary black hole simulation at most two or
three ``orbits'' before coalescence. At these late times the binary
evolution is far from adiabatic, spacetime is highly dynamical, the
separation of the black holes is small compared to the system's total
mass, and it is meaningless to talk of separate black hole masses and
spins, distinguish between total and orbital angular momentum, or
orbital energy, total mass and radiation.

In fact, the situation is not as grim as one might suppose. In fact,
what we are interested in is not some impossible definition of
component masses and spins during the system's penultimate orbit;
rather, it is the masses, spins, and orbital characteristics of the
data's {\em astrophysical antecedent:} the system that evolved from
large separation, where component masses and spins, and orbital energy
and angular momentum, have sensible cognates, to the very compact
state that is the initial data of our binary black hole coalescence
calculation.

One approach to this problem, being explored by Kidder and Finn at
Northwestern University, is to evolve binary systems, using
post-Newtonian perturbation techniques, from wide to compact
separation and compare, in a common gauge, the results of these
perturbation calculations with conventional binary black hole initial
data sets~\cite{cook93a}. In this way perturbation calculations can be
used to characterize, at least in a rough sense, the antecedents of
binary black hole initial data sets.

\section{Conclusions}

The nature of binary black hole coalescence is the final, uncharted
frontier of the relativistic Kepler problem. In the United States,
binary black hole coalescence has been identified as a computational
``Grand Challenge'' whose solution is the object of a coordinated
effort, just reaching its half-way point, by more than two-score
researchers at nearly a dozen institutions.

The computational burden associated with binary black hole coalescence
can be estimated. Using current numerical algorithms a simulation of
binary black hole coalsecence, beginning one or two orbits before
coalescence and ending when the final black hole is weakly perturbed,
requires terabyte computer primary memory and multi-terabyte secondary
storage. In order that such a simulation take less than 12~hours of
computing time, the computer must be capable of a sustanined
performance of $\gtrsim10^12$~floating point operations per second. In
both memory and time these requirements exceed the capabilities of
present day computers by several orders of magnitude and will stretch
the capabilities of the next generation. If we are to be able to solve
for the coalescence of a black hole binary numerically new and more
efficient computational methods must be developed for this problem.

Even with faster computers and new, more efficient computational
methods, there are still open questions of physics associated with the
problems formulation remain to be settled: Can black holes be excised
from the computational domain by exploiting the apparent horizon as a
Cauchy horizon? Can perturbative methods be used to extract radiation
and pose out-going wave boundary conditions at finite distance from
the binary system, or must an interior Cauchy evolution be matched to
a characteristic surface exterior evolution that extends to future
null infinity? How do we relate two black holes, deep in their common
potential and no more than a few orbits prior to coalescence, to a
binary system with well-defined component masses and spins and orbital
energy and angular momentum? The questions of physics raised by the
goal of completing the solution to the relativistic Kepler problem
make the journey itself as interesting and exciting as the
destination.

\section*{Acknowledgments}

It is a pleasure to acknowledge the support of the Alfred P. Sloan
Foundation and the National Science Foundation (PHY/ASC93-18152, ARPA
supplemented).


\begin{thebibliography}{10}

\bibitem{arnowitt62a}
R.~Arnowitt, S.~Deser, and C.~W. Misner.
\newblock The dynamics of general relativity.
\newblock In L.~Witten, editor, {\em Gravitation}, pages 227--265. Wiley, New
  York, 1962.

\bibitem{york79a}
James~W. York, Jr.
\newblock Kinematics and dynamics of general relativity.
\newblock In Larry~L. Smarr, editor, {\em Sources of Gravitational Radiation},
  pages 83--126. Cambridge University Press, Cambridge, 1979.

\bibitem{abrahams90a}
Andrew~M. Abrahams and Charles~R. Evans.
\newblock Gauge-invariant treatment of gravitational radiation near the source:
  analysis and numerical simulations.
\newblock {\em Phys. Rev. D}, 42(8):2585--2594, 15 October 1990.

\bibitem{anninos95c}
Peter Anninos, Karen Camarda, Joan Masso, Edward Seidel, Wai-Mo Suen, and John
  Towns.
\newblock Three dimensional numerical relativity: the evolution of black holes.
\newblock {\em Phys. Rev. D}, 52(4):2059--2082, 15 August 1995.

\bibitem{dongarra94a}
J.~J. Dongarra.
\newblock Performance of various computers using standard linear equations
  software.
\newblock Technical Report CS-89-85, Computer Science Department, University of
  Tennessee, 1994.

\bibitem{gottleib77a}
D.~Gottleib and S.~A. Orszag.
\newblock {\em Numerical Analysis of Spectral Methods: Theory and
  Applications}.
\newblock SIAM, Philadelphia, Pennsylvania, 1977.

\bibitem{bernstein94a}
David Bernstein, David Hobill, Edward Seidel, and Larry Smarr.
\newblock Initial data for the black hole plus {B}rill spacetime.
\newblock {\em Phys. Rev. D}, 50(6):3760--3782, 15 September 1994.

\bibitem{kidder-p1}
Lawrence~E. Kidder and Lee~Samuel Finn.
\newblock Spectral methods for numerical relativity: The initial value problem.
\newblock in preparation, 1996.

\bibitem{abrahams88a}
Andrew~M. Abrahams and Charles~R. Evans.
\newblock Reading off gravitational radiation waveforms in numerical relativity
  calculations: {M}atching to linearized gravity.
\newblock {\em Phys. Rev. D}, 37(2):318--332, 15 January 1988.

\bibitem{price94a}
Richard~H. Price and Jorge Pullin.
\newblock Colliding black holes: the close limit.
\newblock {\em Phys. Rev. Lett.}, 72(21):3297--3300, 23 May 1994.

\bibitem{abrahams95a}
Andrew~M. Abrahams, Stuart~L. Shapiro, and Saul~A. Teukolsky.
\newblock Calculations of gravitational wave forms from black hole collisions
  and disk collapse: Applying perturbation theory to numerical spacetimes.
\newblock {\em Phys. Rev. D}, 51(8):4295--4301, 15 April 1995.

\bibitem{clarke95a}
Chris J.~S. Clarke, Ray~A. d'{I}nverno, and James~A. Vickers.
\newblock Combining {C}auchy and characteristic codes. {I}. the vacuum
  cylindrically symmetric problem.
\newblock {\em Phys. Rev. D}, 52(12):6863--6867, 15 December 1995.

\bibitem{dubal95a}
Mark~R. Dubal, Ray~A. d'{I}nverno, and Chris J.~S. Clarke.
\newblock Combining {C}auchy and characteristic codes. {II}. the interface
  problem for vacuum cylindrical symmetry.
\newblock {\em Phys. Rev. D}, 52(12):6868--6881, 15 December 1995.

\bibitem{seidel92b}
Edward Seidel and W.~Suen.
\newblock Towards a singularity-proof scheme in numerical relativity.
\newblock {\em Phys. Rev. Lett.}, 69:1845--1848, 28 September 1992.

\bibitem{anninos95b}
Peter Anninos, Greg Daues, Joan Masso, Edward Seidel, and Wai-Mo Suen.
\newblock Horizon boundary conditions for black hole spacetimes.
\newblock {\em Phys. Rev. D}, 51(10):5562--5578, 15 May 1995.

\bibitem{scheel95a}
Mark~A. Scheel, Stuart~L. Shapiro, and Saul~A. Teukolsky.
\newblock Collapse to black holes in {B}rans-{D}icke theory. {I}. horizon
  boundary conditions for dynamical spacetimes.
\newblock {\em Phys. Rev. D}, 51(8):4208--4235, 15 April 1995.

\bibitem{bona95a}
Carles Bona, Joan Mass{\'o}, Edward Seidel, and Joan Stela.
\newblock New formalism for numerical relativity.
\newblock {\em Phys. Rev. Lett.}, 74(4):600--603, 24 July 1995.

\bibitem{abrahams95b}
A.~Abrahams, A.~Anderson, Y.~Choquet-Bruhat, and J.~W. York, Jr.
\newblock Einstein and {Y}ang-{M}ills theories in hyperbolic form without
  gauge-fixing.
\newblock {\em Phys. Rev. Lett.}, 75(19):3377--3381, 6 November 1995.

\bibitem{vanputten95Pa}
Maurice H. P.~M. van Putten and Douglas Eardley.
\newblock Nonlinear wave equations for relativity.
\newblock preprint available as gr-qc/9505023, May 1995.

\bibitem{bruhat95Pa}
Y.~Choquet-Bruhat and J.~W. York, Jr.
\newblock Geometrical well posed systems for the {E}instein {E}quations.
\newblock preprint available as gr-qc/9506071, June 1995.

\bibitem{cook93a}
Gregory~B. Cook, Matthew~W. Choptuik, Mark~R. Dubal, Scott Klasky, Richard~A.
  Matzner, and Samuel~R. Oliveira.
\newblock Three dimensional initial data for the collision of two black holes.
\newblock {\em Phys. Rev. D}, 47(4):1471--1490, February 1993.

\end{thebibliography}

\end{document}